\documentclass[aps,pra,preprint,superscriptaddress,showpacs]{revtex4}

\usepackage{amsmath,amssymb,amsfonts}
\usepackage{graphicx}

\newcommand{\ket}[1]{\vert{#1}\rangle}
\newcommand{\fref}[1]{Fig.~\ref{#1}}

\newcommand{\eqnref}[1]{Eq.~(\ref{#1})}

\begin{document}

\title{Effect of Squeezing on the Atomic and the Entanglement Dynamics in the Jaynes-Cummings Model}

\author{T. Subeesh}
\affiliation{Department of Physics, Indian Institute of Technology Madras, Chennai 600036, India}
\affiliation{Department of Physics, Amrita School of Engineering, Amrita Vishwa Vidyapeetham, Ettimadai, Coimbatore 641105, India}
\email[]{t.subeesh@gmail.com}

\author{Vivishek Sudhir}

\affiliation{Department of Physics, Imperial College, London, SW7 2AZ, UK}

\author{A. B. M. Ahmed}
\affiliation{Department of Physics, Indian Institute of Technology Madras, Chennai 600036, India}
\affiliation{School of Physics, Madurai Kamaraj University, Madurai 625 021, India}

\author{M. Venkata Satyanarayana}
\affiliation{Department of Physics, Indian Institute of Technology Madras, Chennai 600036, India}
\email[]{mvs@physics.iitm.ac.in}

\date{\today}

\begin{abstract}

The dynamics of the Jaynes-Cummings interaction of a two-level atom interacting with a single  mode of the radiation field
is investigated, as the state of the field is gradually changed from a coherent  state to a squeezed coherent state. The effect of mild squeezing on the coherent light is shown to be striking:
the photon number distribution gets localized and it peaks maximally for a particular value of squeezing. The atomic inversion retains its
structure for a longer time. The mean linear entropy shows  that the atom has a  tendency to get disentangled from field within the collapse region
and also in the revival region, for mild squeezing. These properties are absent for the case of a coherent state or for an excessively  squeezed coherent state.
We also elucidate a connection between these properties and the photon statistics of the mildly squeezed coherent state; these states have the minimum variance
and are also maximally sub-Poissonian.

\end{abstract}

% insert suggested keywords - APS authors don't need to do this
%\keywords{}

\maketitle

\section{Introduction}

The squeezed coherent states are very well known. It is also well known that the photon-number distribution, $P(n)$,
associated with certain squeezed coherent states exhibits dramatic oscillations and the oscillatory behavior can be interpreted using
the concept of phase space interference \cite{SchWhe87Jopt,SchWhe87Nat}.
The interaction of a coherent state of the single mode field with a two-level atom has been studied
via the Jaynes-Cummings model \cite{Jay63,Cum65,KniRad82,ShoKni93}. Interesting features of the atomic inversion and the granular nature
of the radiation field have been predicted and experimentally realized \cite{Rem87}. Other than being the paradigm model of optical resonance
describing the interaction of a single two-level atom with a single mode of radiation field, the Jaynes-Cummings model has also been
shown to be of tremendous relevance in the possibility of realizing a quantum computer using cold trapped ions \cite{CirZol95}.

In this paper, we consider the
Jaynes-Cummings interaction of a single two-level atom interacting with a single mode of radiation field prepared in a
coherent state and the field statistics is changed from the coherent state to a squeezed coherent state by gradually increasing the squeezing.
Milburn has studied the interaction of a two-level atom and a single mode of radiation field with the field prepared in a squeezed
coherent state and he has shown that the atomic response is similar to that of chaotic radiation for those states for which
the coherent contribution to the photon-number variance is dominant \cite{Mil84}. In another contrasting study, the response of an
atom, where the squeezed contribution to the photon number variance is dominant, has been investigated \cite{Mvs89}.
This situation was very specifically chosen in order to study the atomic inversion when the photon-counting
distribution is sharply oscillatory.

The objective of this paper is to study the photon-counting distribution, the atomic inversion and the entanglement
dynamics corresponding to the case in which the squeezed contribution to the mean number of photon is small
compared to the coherent contribution to the mean number of photons. This will facilitate us to start with a coherent
state and increase the squeezing in such a way that the granular effect of squeezing on the photon-counting distribution,
the atomic inversion and the entanglement dynamics can be studied. Further, we show that in this regime of
squeezing, there exists an ``optimal'' level of squeezed contribution  that is able to sustain the minimal atom-field entanglement
for a longer period of time. We find that this particular condition of squeezing is intimately tied with the sub-Poisson statistics
of its photon number distribution.

\section{The photon number distribution and atomic dynamics}

The interaction of a two-level atom with a single mode of a radiation field given by Jaynes and Cummings is
\begin{eqnarray}
\label{JCH}
\hat{H}=\hbar \omega_0 \left( \hat{a}^{\dagger}\hat{a}+\frac{1}{2}\right) + \frac{\hbar \omega_0}{2}\; \hat{\sigma}_z
+\hbar \lambda \left( \hat{\sigma}_+ \hat{a}+\hat{\sigma}_- \hat{a}^{\dagger}\right),
\end{eqnarray}
where $\hat{\sigma}_+$, $\hat{\sigma}_-$ and $\hat{\sigma}_z$ are the Pauli pseudo-spin operators; $\hat{a}$ and
$\hat{a}^{\dagger}$ are the photon annihilation and creation operators; $\lambda$ is the coupling constant describing
the atom-field interaction; and $\omega_0$ is the frequency of the radiation-field mode. The
atomic population inversion, when the atom is initially prepared in its ground state, is
\begin{eqnarray}
\label{wt}
W(t)=-\frac{1}{2} \sum_{n=0}^{\infty} P(n) \cos(2 \lambda t \sqrt{n}),
\end{eqnarray}
where $P(n)$ is the photon-number distribution of the initial state of the radiation field.

For the initial state of the radiation field, we consider the squeezed coherent state,
\begin{eqnarray}
\label{SCS}
\ket{\alpha, z} = \hat{D}(\alpha) \hat{S}(z) \ket{0},
\end{eqnarray}
where $\hat{S}$ is the squeezing operator
\begin{eqnarray}
\label{SO}
\hat{S}(z)=\exp \left[ \frac{1}{2} \left( z^* \hat{a}^2 - z \hat{a}^{\dagger^2} \right) \right],
\end{eqnarray}
and $\hat{D}(\alpha)$ is the displacement operator.
\begin{eqnarray}
\label{DO}
\hat{D}(\alpha)=\exp \left( \alpha \hat{a}^{\dagger}-\alpha^* \hat{a} \right).
\end{eqnarray}
The photon-number distribution for the squeezed coherent state is given by \cite{Yue76, Lou87},
\begin{eqnarray}
\label{pnSCS}
P(n)=\frac{1}{n! \mu} \left( \frac{\nu}{2 \mu}\right)^n H_n^2 \left(\frac{\beta}{\sqrt{2\mu \nu}}\right)
\exp \left[-\beta^2 \left( 1-\frac{\nu}{\mu} \right) \right],
\end{eqnarray}
where $\mu=\cosh |z|$ and $\nu=\sinh |z|$ and $\beta = |\alpha| (\mu+\nu)$. The average number of photons in a squeezed
coherent state is given by,
\begin{eqnarray}
\label{N_SCS}
\langle n \rangle = N_{SCS}= N_C + N_S,
\end{eqnarray}
where $ N_C = |\alpha|^2 $ and $N_S=\vert \nu \vert^2 = \sinh^2 |z|$. $N_C$ and $N_S$ are respectively the coherent contribution and the squeezed
contribution to $N_{SCS}$. $P(n)$ tends towards the photon-number distributions for pure squeezed radiation and coherent
radiation, respectively, as $N_C \rightarrow 0$ and $N_S \rightarrow 0$.

In view of the fact that $P(n)$ in one extreme limit describes a coherent state and in another extreme limit describes a
pure squeezed state, the study of atomic inversion of a two level atom interacting with a single mode of electromagnetic
field in a squeezed coherent state enables us to observe how the revivals of atomic inversion in a coherent state
develop into seemingly chaotic oscillation and vice versa. Also, this approach is useful to study the role of squeezing on
the entanglement dynamics.

\begin{figure}[h!]
\centering
\includegraphics[scale=0.75]{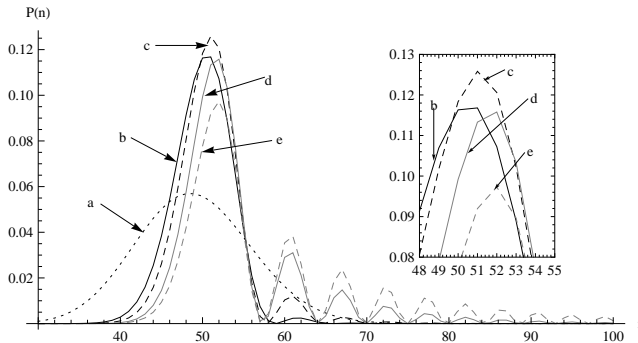}
    \caption{$P(n)$ versus $n$ for squeezed coherent state with $N_{C} = 49$ for different squeezing parameters. (a) coherent state,
            (b) $N_{S}=1$, (c) $N_{S}=2$, (d) $N_{S}=5$, (e) $N_{S}=10$.
    \label{fig1.eps}}
\end{figure}

\begin{figure}[h!]
\centering
\includegraphics[scale=0.8]{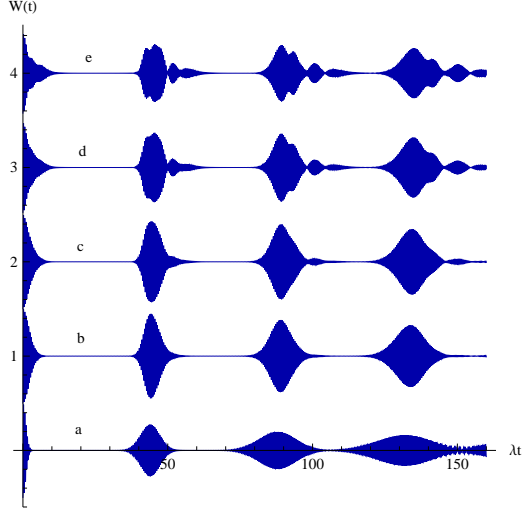}
    \caption{$W(t)$ versus $\lambda t$ for squeezed coherent state with $N_{C} = 49$ for different squeezing parameters. (a) coherent state,
        (b) $N_{S}=1$, (c) $N_{S}=2$, (d) $N_{S}=5$, (e) $N_{S}=10$.
    \label{fig2.eps}}
\end{figure}

Our calculations show that the effect of squeezing on coherent light is very dramatically manifested in the photon-counting
distribution. An important consequence is the localization of the photon-counting distribution.
In \fref{fig1.eps}, we plot the photon-number distribution $P(n)$ as given by \eqnref{pnSCS} for the case in which the
mean number of coherent photons $N_C=49$ and $N_S$ runs over the values $0, 1, 2, 5$ and $10$. Curves \textit{b} and \textit{c} reveal that
even if $N_S = 1$ or $2$ (when the coherent contribution to the mean is 49), the shape of the distribution is significantly different from
that of a coherent state (curve \textit{a}). The peak of the distribution rises by about $100\%$ (or more),
and the distributions narrow simultaneously. We shall call this as the `localization of photon-counting distribution'. The
oscillations in the distributions show up as mild but visible ripples. As $N_S$ is increased further ($N_S=5,10$), the peak of the distributions
fall slightly, and the secondary oscillations become more pronounced.

\begin{figure}[h!]
\centering
\includegraphics[scale=0.75]{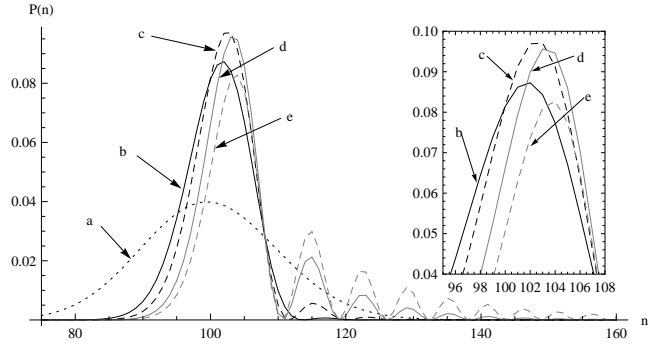}
    \caption{$P(n)$ versus $n$ for squeezed coherent state with $N_{C}= 100$ for different squeezing parameters. (a) coherent state,
        (b) $N_{S}= 1$, (c) $N_{S}= 2$, (d) $N_{S}= 5$, (e) $N_{S}= 10$.
    \label{fig3}}
\end{figure}

This remarkable sensitivity of the distribution to the
number of squeezed photons added to the coherent field is further highlighted in \fref{fig4}, which corresponds to the mean coherent photon number
$N_C=100$ and $N_S=0, 1, 2, 5$ and $10$. Curve \textit{b} illustrates very prominently the localization of the photon-counting distribution due to
squeezing. Furthermore,  the rise of the peak in curve \textit{c} of \fref{fig1.eps}, in
comparison with the peak height of curve \textit{b} in \fref{fig1.eps}, is not as dramatic as the rise of the peak in curve \textit{b} of
\fref{fig1.eps}, in comparison with the peak height of curve \textit{a} in \fref{fig1.eps}. Squeezing further, the peak of the distributions
begin to fall; the distributions in fact become short and narrow and pick up oscillations. A similar trend is shown by squeezed coherent state with $N_C = 100$
as depicted in \fref{fig4}.
Thus, an initial squeezing of a coherent state, even by very small amounts, is sufficient to localize the distribution; but further squeezing results in
the building up of oscillations than to rise the peak.

The consequences of such dramatic changes in the photon number distributions of the radiation field manifests itself
on the atomic population inversion of a two-level atom, as brought out in \fref{fig2.eps}.
Curve \textit{a}, in  \fref{fig2.eps}  depicts the familiar collapse and revival phenomena associated
with a two-level atom interacting with a coherent state, whose photon number distributions is given in curve \textit{a} of
\fref{fig1.eps}. The curves \textit{d} and \textit{e} in \fref{fig2.eps}  depict the remarkably different
collapse and ``ringing'' revivals \cite{Mvs89} as higher number of squeezed photons are added into the coherent field.

The differences between curves \textit{a} and \textit{b} of \fref{fig2.eps} are amply evident. The widths of the revival-collapse structure have
shrunk considerably; and, at this stage, collapses following the revivals do not ``ring'' yet. This is expected since
the $P(n)$ represented by the curve \textit{b} in \fref{fig1.eps} does not have dominant oscillations. An important fact that
emerges is that the collapse times in \textit{b} are longer than the corresponding collapse times in \textit{a}. In other words, the process
of revival in atomic inversion is delayed by an increase in $N_S$. The ``ringing'' structures in collapses gradually appear
as $N_S$ increases, as evident from the curves \textit{c}, \textit{d} and \textit{e} in \fref{fig2.eps} -- a reflection of the fact that the
photon-counting distribution curves \textit{c}, \textit{d} and \textit{e} in \fref{fig1.eps} display increasing oscillatory nature in that order.
So, the $W(t)$ begins to exhibit ringing structures only for those distributions which are significantly oscillatory.

The generic squeezed coherent state is thus an ideal state of the radiation field for the understanding of the transition
from the regular to seemingly irregular dynamics of atomic inversion. We may begin with a radiation field prepared in the
coherent state ($N_C > 0$ and $N_S=0$), generate a sequence of squeezed coherent states by increasing $N_S$, and study the
dynamics of atomic inversion during this process. On the other hand, if we begin with a radiation field prepared in
squeezed vacuum ($N_C=0$ and $N_S>0$), and generate a sequence of squeezed coherent states by increasing $N_C$, it would
be of little consequence to the resulting photon-counting distributions and also to the behavior of $W(t)$ -- it would
continue to remain similar to that of chaotic radiation.

It is of interest to compare the effect of squeezing on a coherent state with those of Glauber-Lachs version of the
Jaynes-Cummings interaction \cite{Mvs92}. An important difference is that a very small amount of squeezing a coherent state ($N_S/N_C \approx 0.01$)
is sufficient to localize the photon-counting distribution very sharply and increase the height of its peaks. The behavior of the photon-counting distributions
of Glauber-Lachs states are quite contrary -- the presence of even one mean thermal photon among $100$ mean coherent photons results in the peak of the distribution
falling by $50\%$ or more and at the same time, causing the distribution to broaden.

\section{Dynamics of Atom-Field Entanglement}

Now, we proceed to study the dynamics of entanglement between the atom and the radiation field produced by the Jaynes-Cummings interaction.
The combined state of the atom and the radiation field at any time $t$ is a pure state and it is given by,
\begin{equation}
\ket{\psi(t)} = \sum_{n}\left( C_{e,n}(t)\ket{e,n}  + C_{g,n}(t)\ket{g,n} \right),
\end{equation}
where $C_{e,n}(t)$ and $C_{g,n}(t)$ are the time-dependent superposition coefficients. In the case of pure states all entanglement measures, such as von Neumann entropy and linear entropy are equivalent \cite{elorany:2007}. Linear entropy is employed in this paper as the entanglement measure as it is easier to compute. Linear entropy is defined as,
\begin{eqnarray}
\label{LE}
L(t)=2\left[ 1-\text{Tr} \left( \hat{\rho}_A^2(t) \right) \right] = 2\left[ 1-\text{Tr} \left( \hat{\rho}_F^2(t) \right)\right],
\end{eqnarray}
where $\hat{\rho}_A(t)$ and $\hat{\rho}_F(t)$ are the reduced density operators for the atom and radiation field respectively at any time $t$. At maximal entanglement between the atom and the radiation field, $L(t) = 1$ whereas $L(t) = 0 $ indicates that the atom and the radiation field are completely disentangled. The reduced density
operator of the atom is,
\begin{equation}
\hat{\rho}_{A}(t) =
\begin{pmatrix}
    \sum_{n}|C_{e,n}(t)|^{2} & \sum_{n}C^{\ast}_{e,n}(t)C_{g,n}(t) \\
    \sum_{n}C_{e,n}(t)C_{g,n}^{\ast}(t) & \sum_{n}|C_{g,n}(t)|^{2}
\end{pmatrix}.
\end{equation}

\begin{figure}[h!]
\centering
\includegraphics[scale=0.8]{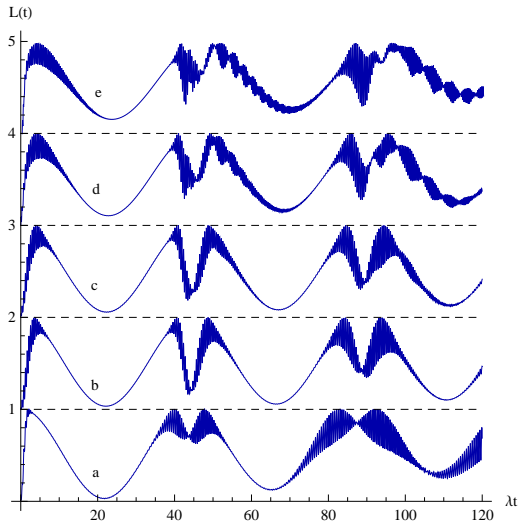}
\caption{Linear entropy versus $\lambda t$ for squeezed coherent state with $N_{C}=49 $ for different squeezing parameters. (a) coherent state,
(b) $N_{S}= 1$, (c) $N_{S}= 2$, (d) $N_{S}= 5$, (e) $N_{S}= 10$}
 \label{fig4}
\end{figure}

Our objective in this section is to study the effect of squeezing on the dynamics of linear entropy. \fref{fig4} depicts the
behavior of linear entropy as a function of scaled time $\lambda t $.
Curve \textit{a} of \fref{fig4}, depicts $L(t)$ for $N_C=49$ and $N_S=0$, corresponding to the distribution shown
in curve \textit{a} of \fref{fig1.eps}, \textit{i.e.}, for a coherent field with absolutely no squeezed photons.
Comparing the time scales of this curve with curve \textit{a} of \fref{fig2.eps}, it is seen that at times which are
half way into the collapse region, the atom and the field have a tendency to disentangle, \textit{i.e.} the atom returns, almost exactly, to a pure state.
For the particular case of a coherent field, this phenomena was noted by Gea-Banacloche \cite{Ban90}.
As time progresses, these nearly exact regenerations of the atom to a pure state continues to a lesser extent; at sufficiently long times, as seen
in curve \textit{a} of \fref{fig6}, the atom completely loses its ability to return to a pure state.

It is most striking that as squeezing is increased, in relatively small numbers, the subsequent minima
(at around $\lambda t = 65$) in the linear entropy are much lower than those for the case of a coherent state; this is clearly seen in curves \textit{b}
and \textit{c} of \fref{fig4}. But as the squeezing is continued further, this property is lost, as seen in curves \textit{d} and
\textit{e} of \fref{fig4}. What is interesting in these latter cases is that the ``ringing'' phenomena seen in the population inversion
curves \textit{d} and \textit{e} of \fref{fig2.eps}, is manifested in the linear entropies as well, particularly in the regions immediately
following the revivals, \textit{i.e.} at around $\lambda t = 55$ and $\lambda t = 100$.

Apart from the ability of the squeezed states, corresponding to curves \textit{b} and \textit{c}, to bring the
atom to a nearly pure state halfway into the collapse region, they also lead to  a remarkable disentanglement of the atom
during revivals. As seen from curves \textit{b} and \textit{c} in \fref{fig4}, during revival, \textit{i.e.} for $\lambda t \sim 45, 85$ etc,
these critical levels of squeezing maximally lowers the entanglement between the atom and the field. For higher squeezing, corresponding to
curves \textit{d} and \textit{e}, these minima at revival raise up, \textit{i.e.}, the atom and the field get more entangled.

\begin{figure}[h!]
\includegraphics[scale=0.8]{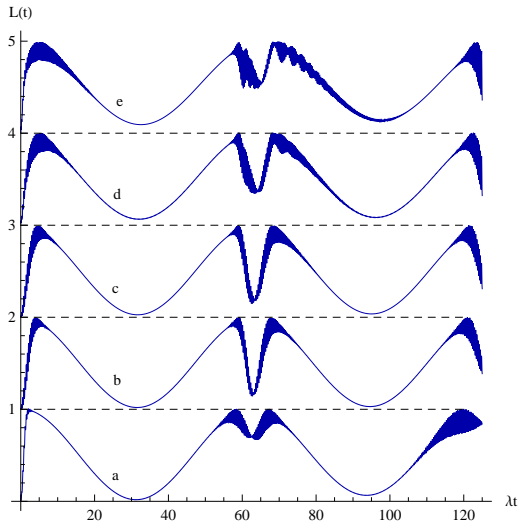}
\centering
    \caption{Linear entropy versus $\lambda t$ for squeezed coherent state with $N_{C} = 100$ for different squeezing parameters. (a) coherent state,
        (b) $N_{S}=1$, (c) $N_{S}=2$, (d) $N_{S}=5$, (e) $N_{S}=10$.}
    \label{fig5}
\end{figure}

The appearance of the inverted low entanglement spikes is further illustrated in \fref{fig5} which corresponds to a mean coherent photon number $N_{C}=100$ and squeezed photon numbers $N_{S}=0,1,2,5,10$.
For a coherent state, \textit{i.e.} curve \textit{a} of \fref{fig5}, the duration of the collapse intervals is larger; this is due to the stronger
mean cavity photon number.
When a small amount of squeezing ($N_{S}=1$) is affected (curve \textit{b}), the entanglement pattern mimics the pattern of curve \textit{a} for the initial rise and fall. However, the difference in the entanglement evolution due to coherent state and squeezed coherent state occurs after this initial cycle -- inverted low entanglement spikes appear half way into the collapse region even for the $1\%$ contribution from the squeezing. Though these features are present in the \fref{fig4}, it is inferred from the \fref{fig5} that the effect of the added squeezed noise is striking even for a stronger coherent field. Curves \textit{c}, \textit{d} and \textit{e} of
\fref{fig5} show the effect of increased squeezing noise in entanglement dynamics, and qualitatively shows the same features observed for the case $N_C = 49$
in \fref{fig4}.

\begin{figure}[h!]
\includegraphics[scale=0.8]{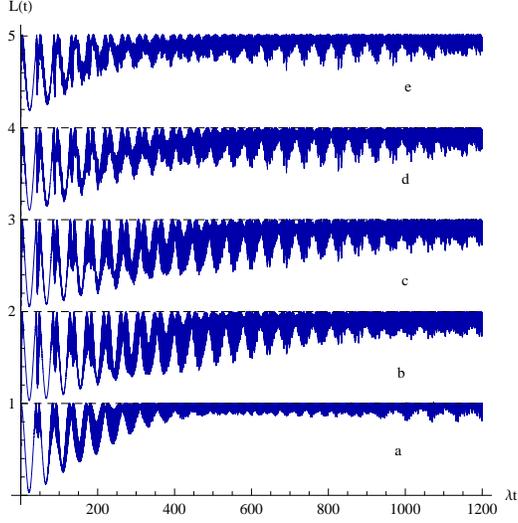}
\centering
\caption{Linear entropy versus $\lambda t$ for squeezed coherent state with $N_{C} = 49$ for different squeezing parameters. (a) coherent state,
(b) $N_{S}=1$, (c) $N_{S}=2$, (d) $N_{S}=5$, (e) $N_{S}=10$.}
 \label{fig6}
\end{figure}

Investigating the long time entanglement behavior for $N_C = 49$, it is seen that the squeezed states corresponding to curves \textit{b} and \textit{c} in
\fref{fig6}, tend to preserve the ability of the atom to resist getting completely entangled with the field. This peculiar behavior of the system, is
directly related to the statistics of the field for that particular amount of squeezed photon addition.

The variance in the photon number distribution of the squeezed coherent state $\ket{\alpha,z}$ can be expressed as \cite{Mil84, Lou87},
\begin{equation}
    \langle ( \Delta n )^2 \rangle = 2 N_S (1+N_S) + N_C \left( 1+ 2N_S -2 \sqrt{N_S (1+N_S)} \right)
\end{equation}
whenever the parameters $\alpha$ and $z$ are taken to be real. Although the variance, and the mean given by \eqnref{N_SCS}, do not completely determine the
photon distribution for $N_S \neq 0$, the former characterizes the spread of the distribution. On the other hand,
Mandel's Q parameter, is defined by,
\begin{equation}
    Q = \frac{\langle ( \Delta n )^2 \rangle}{\langle n \rangle} - 1;
\end{equation}
when $Q=0$, one has Poisson statistics, akin to that of a coherent state, and for $Q < 0$ ($Q > 0$), one has sub(super)-Poisson statistics.

\begin{figure}[h!]
\centering
\includegraphics[scale=0.7]{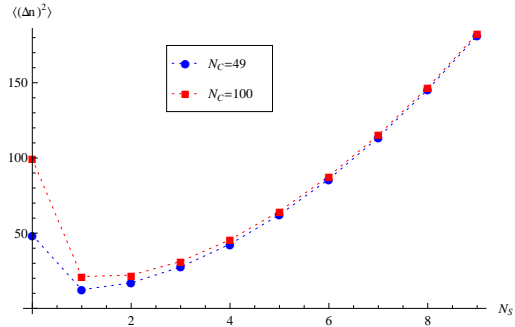}
    \caption{$\langle ( \Delta n )^2 \rangle$ versus $N_S$ for squeezed coherent state with $N_{C}= 49, 100$.
    \label{fig7}}
\end{figure}

\begin{figure}[h!]
\centering
\includegraphics[scale=0.7]{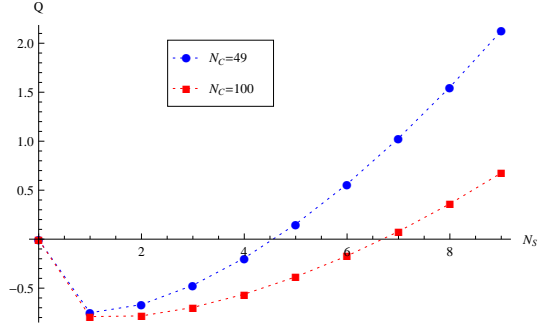}
    \caption{Mandel's Q parameter versus $N_S$ for squeezed coherent state with $N_{C}= 49, 100$.
    \label{fig8}}
\end{figure}

\fref{fig7} and \fref{fig8} depict respectively, the variance in the photon number distribution and Mandel's Q parameter for various levels of squeezing
when $N_C = 49$ and $N_C = 100$. It is seen that for both $N_C =49$ and $N_C = 100$, there exists a few cases where the  squeezed
contribution decreases the variance of the photon number distribution. It is precisely these critical number of mean squeezed photons that leads to the
maximum localization of the photon counting distribution mentioned earlier. These are also the states that show a minima in the Mandel Q parameter in the sub-Poisson regime;
further squeezing results in super-Poisson statistics. If a squeezed coherent state has a total of $N_{SCS}$ mean  photons, of which $N_C$ are the mean coherent photons, then for the state to be
a `minimum Q state', it is required that the mean number of squeezed photons in it, $N_S$, to satisfy,
\begin{equation}\label{Ns_minQ}
    2N_{SCS}^2 + N_C \left( 1 + (1+ N_S^{-1})^{-1/2} \right) = N_C^2 (1 + N_S^{-1})^{1/2}.
\end{equation}
On the other hand, states which are maximally localized are the ones that have minimum variance in its photon number distribution; these states are the ones whose $N_S$
satisfies the equation,
\begin{equation}\label{NS_minVarN}
    2 (1+N_C+2N_S) \sqrt{N_S (1+N_S)} = N_C (1+2N_S).
\end{equation}

\begin{figure}[h!]
\centering
    \includegraphics[scale=0.75]{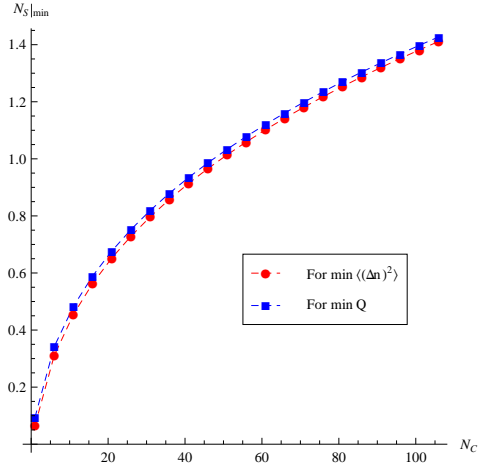}
    \caption{Plot of the numerical solution of \eqnref{Ns_minQ} (shown using asterisk) and \eqnref{NS_minVarN} (shown using +) in the range $ 1 \leq N_C \leq 100$.
    \label{fig9}}
\end{figure}

\fref{fig9} shows the solutions, $N_S$, of \eqnref{Ns_minQ} and \eqnref{NS_minVarN} for $1 \leq N_C \leq 100$. It is evident that the `minimum Q' squeezed
states are the ones with minimum photon number variance and are hence the ones with the highest localization in the number distribution. The solution points of
\fref{fig9} represent the average numbers of squeezed photons to be added to the Poissonian coherent field to make the resulting field maximally
sub-Poissonian. It is obvious from this plot that the number of squeezed photons that results in a `minimum Q state' for the cases $N_C = 49$ and $N_C = 100$,
are the same ones corresponding to curves \textit{b} in \fref{fig2.eps}, \fref{fig4}, \fref{fig5} and \fref{fig6}. Also,
as noted in preceding paragraphs, these same curves, \textit{b}, are the ones where interesting phenomena like -- regeneration of the atomic pure state in the
collapse region, inverted low entanglement spikes in the revival regions, and tendency to preserve the atom's ability to resist getting completely mixed with the field over
long time scales -- are all significantly pronounced and drastic.

The connection between the `minimum Q' squeezed coherent state and the entanglement properties of the atom can be succinctly captured by the mean entanglement between
the atom and the field; the ideal measure of this is the time average of the linear entropy,
\begin{equation*}
    \bar{L} = \frac{1}{T} \int_0^T L(t) dt,
\end{equation*}
where $L(t)$ is given by \eqnref{LE}.

\begin{figure}[h!]
\centering
\includegraphics[scale=0.75]{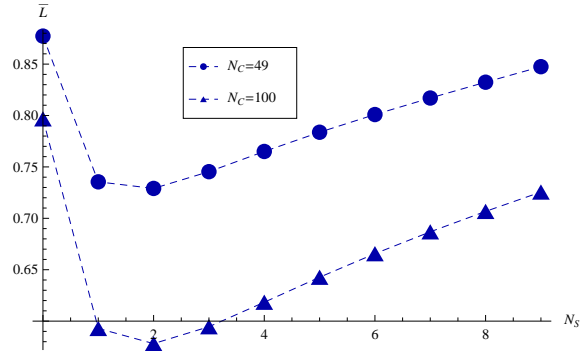}
\caption{Time averaged linear entropy plotted against the number of squeezed photons added. The mean is calculated over time scales $\lambda T = 1000$.}
\label{fig10}
\end{figure}

\fref{fig10} shows the time averaged linear entropy, and it is clearly visible that the minimum mean entropy occurs at the same value of $N_S$ for which the
Mandel Q parameter and the variance in the photon number distribution are also minimum. This means that the atom and field remain minimally entangled for a longer duration of
time when this critical number of mean squeezed photons are added to the coherent field.

\section{Conclusion}
Squeezing a coherent state very mildly effects in a strong localization of the photon counting distribution $P(n)$. Further squeezing is responsible for the oscillations in the $P(n)$.  
 The effect of squeezing in the dynamics of  Jaynes-Cummings interaction  has been
investigated.
The system shows remarkable features for an optimum amount of squeezing. At optimum squeezing, the photon number distribution has minimum variance and is sub-Poissonian. The atomic population inversion is found to remain in the collapse region for a longer time. It is also seen that the atom remains less mixed, thus minimally entangled for a longer period during its evolution in comparison with the  coherent state.

\begin{acknowledgements}
The authors thank Professors Arul Lakshminarayan, P. C. Deshmukh and S. Sivakumar for the discussions and suggestions.
\end{acknowledgements}

\bibliography{refpap3}

\end{document}